# Algebraic algorithm for next-to-leading order calculations in the large-*s*/small-*x* regime

A.V. Stepanova[a] and F.V. Tkachov[b]

[a]Physics Department, Moscow State University, Moscow 119899
[b]Institute for Nuclear Research of Russian Academy of Sciences, Moscow 117312

An algebraic algorithm is presented for analytical calculation of arbitrary dimensionally regulated massless two-loop forward-scattering diagrams that constitute the most cumbersome part of next-to-leading order calculations in the large-*s*/small-*x* regime.

Introduction                                                    1

There is much interest in studying processes that can be described as almost-forward scattering: e.g. the small-*x* problem in the context of deep-inelastic scattering in QCD [1] (via the optical theorem), the small-angle Bhabha scattering in QED [2], etc. In such problems one deals with a kinematical situation when the scattering of two particles $k_1 + k_2 \to k_3 + k_4$ (e.g. virtual photon and proton in the case of DIS; electron and positron in the case of Bhabha scattering) occurs at the values of the Mandelstam variable $s = (k_1 + k_2)^2$ (corresponding to $W^2$ in the DIS case) much larger than any other Lorentz-invariant dimensional parameter in the problem (masses and virtualities of participating and virtual particles as well as the Mandelstam variables $t$ and $u$). A systematic way to treat such processes in perturbative Quantum Field Theory is via an asymptotic expansion corresponding to the asymptotic regime $s \gg t, u, k_i^2$. This asymptotic regime belongs to the so-called class of non-Euclidean asymptotic regimes for which, unlike the Euclidean class, systematic prescriptions for obtaining asymptotic expansions to all orders of the small expansion parameter (in our case, $s^{-1}$) did not exist (see e.g. [3]). The first systematic theory of non-Euclidean asymptotic regimes based on the distribution-theoretic method of asymptotic operation [3] has been reported in another talk at this workshop [4]. According to the theory of [3], [4], systematic asymptotic expansions are obtained by first formally Taylor-expanding the corresponding diagrams in the small parameter, and then adding certain counterterms whose structure is dictated by the prescriptions of the theory of asymptotic operation. The key point for us here is that those counterterms always contain δ-functions instead of some propagators of the original diagram, so they are easier to take into account than the formally expanded part, although from theoretical point of view it is the obtaining of counterterms that is the most interesting part of the problem. From calculational point of view, however, the most cumbersome part is always the calculation of the formally expanded diagrams of the original process (this is also the case with the NNLO calculations of coefficient functions of OPE — e.g. [5] — according to the prescriptions of the Euclidean theory of asymptotic operation [6], [7]). It is this problem that we are going to address in this work.

Topologies and algorithm                                       2

The loop integrals we are going to consider have the following kinematics:

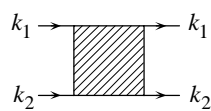

2.1

One-loop diagrams of this type are elementarily integrated in terms of Euler's Γ-functions. So we are going to consider two-loop diagrams. The method we will use is an extension to this kinematics of the well-known algebraic (integration-by-parts) algorithm of [8] that was originally developed for 3-loop massless self-energy diagrams.

There are eighteen non-equivalent two-loop topologies. (For instance, topologies that differ e.g. by $k_1 \to -k_2$ are equivalent.) Here is the list:

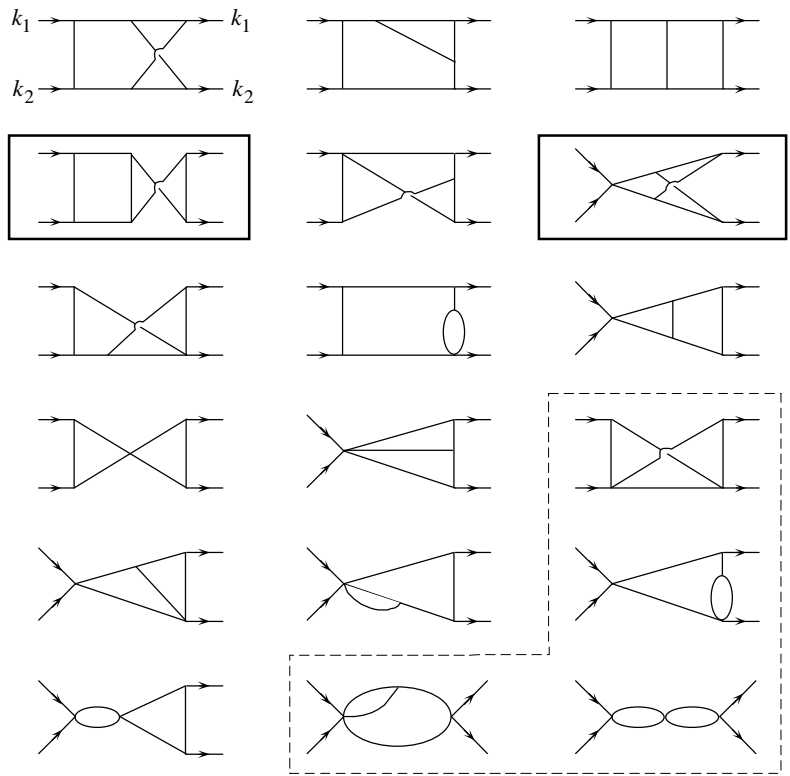

2.2

Four of them (enclosed in the dashed box) can be calculated in Γ-functions by direct integration. Twelve topologies can be treated using integration-by-parts recurrencies. The remaining two (in solid boxes) are irreducible in the sense that they contain some integrals that can neither be reduced to other topologies via integration by parts nor easily expressed in terms of Γ-functions.

The integration-by-parts method [8] is well known and we do not explain it in details. We just present one example of a recursion relation for one of the topologies:

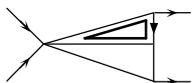

2.3

This in fact is similar to one of the topologies encountered in the original integration-by-parts algorithm for three-loop self energies:

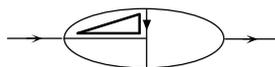

2.4

It is sufficient to apply the rule of triangle [8] to the triangle indicated on Fig. 2.3. But now we have massless external momenta ($k_1^2 = k_2^2 = 0$). Therefore whenever a $k_i^2$ occurs in the identity, the corresponding contribution vanishes whereas in the original algorithm it caused a line to shrink to a point. In particular, the term that corresponded to shrunk top right line in 2.4 is nullified in the case of 2.3.

The resulting recurrence relation has a standard form, and we only point out that the integrals resulting from repeated application of the rule of triangle according to 2.3 all belong to the same topology:

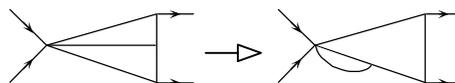

2.5

## Irreducible topologies 3

One class of irreducible diagrams corresponds to the following topology:

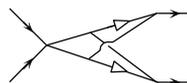

3.1

Note that for this topology, there is an irreducible scalar product which can be chosen as the scalar product of the momenta flowing through the lines shown with hollow arrows. In the original integration-by-parts algorithm it was possible to get rid of the powers of the irreducible scalar product by differentiations with respect to the external momentum. In the present case the two external momenta are restricted to mass shell, so the differentiation does not work. Using symmetry with respect to reflections, one can get rid of odd powers of the irreducible invariant. In practice we expect only zeroth and second powers to occur.

The other irreducible topology contains only one irreducible diagram:

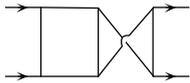 3.2

with all powers of propagators equal to one.

Detailed discussion of calculation of the irreducible topologies goes beyond the scope of the present paper. We only note the following. The diagram 3.2 can be extracted as a special case from the double box result of N.I.Ussyukina reported at this workshop [9], and the topologies 3.1 can be done analytically using similar methods. Alternatively, all irreducible diagrams can be connected with convergent diagrams which can be computed numerically.

## Conclusions 4

In conclusion we compare our algorithm with the original algorithm for massless 3-loop self-energies. First, in the case of 2-loop forward scattering we have more topologies than in the case of 3-loop self-energies. Second, now there are more types of diagrams to be computed using special formulas. Third, there are more irreducible diagrams. Forth, there are more poles here — fourth order poles (two poles per loop as is typical for Minkowski-space integrals) instead of third order poles in the case of the original algorithm. So, on the whole the algorithm is more cumbersome. However, this is not expected to be a problem because we are going to use a specialized software package BEAR[a] which is designed to be much faster and much more economical with disk space than analogous software currently available such as the well-known MINCER program [10], [11] for systems such as Veltman's SCHOONSCHIP [12].

## Acknowledgments

We thank N.I. Ussyukina for discussion of irreducible diagrams, the organizing committee and sponsors of the QFTHEP'97 workshop for financial support, and the participants of the workshop for comments. This work was supported in part by the Russian Foundation for Basic Research under grant 95-02-05794.

---

[a] http://www.inr.ac.ru/~ftkachov/projects/bear/index.htm